\documentstyle[12pt]{article}
\newcommand{\be}{\begin{equation}}
\newcommand{\ee}{\end{equation}}
\newcommand{\benonum}{\begin{equation*}}
\newcommand{\eenonum}{\end{equation*}}
\newcommand{\ba}{\begin{eqnarray}}
\newcommand{\ea}{\end{eqnarray}}
\newcommand{\banonum}{\begin{eqnarray*}}
\newcommand{\eanonum}{\end{eqnarray*}}

\newcommand{\tr}{{\rm tr}}

\newcommand{\half}{{1\over2}}
\newcommand{\quarter}{{1\over4}}
\newcommand{\gij}{{g_{ij}}}

\begin{document}

\begin{flushright}
hep-th/9907053\\
QMW-PH-99-09\\
\end{flushright}

\hskip 1.5cm
\vfill

\begin{center}
{\LARGE Topological Born-Infeld Actions and D-Branes
}\vfill
{\large B. Spence$^*$\\
{\it Department of Physics\\ Queen Mary and 
Westfield College\\
 Mile End Road\\
London E1 4NS UK.}

\vspace{6pt}
}
\end{center}

\vfill

\begin{abstract}
We propose that the effective
field theories of certain wrapped D-branes are
given by topological actions based on Born-Infeld theory.
In particular, we present a Born-Infeld version of the Abelian
Donaldson-Witten theory. We then consider wrapping D3 branes on
calibrated submanifolds and for the Calabi-Yau four-fold case,
discuss how the resulting theory
could give rise to a Born-Infeld version of the ampicheiral twisted
$N=4$ super Yang-Mills topological field theory.
\end{abstract}

 \vfill \hrule width 3.cm
{\footnotesize
\noindent
 $^*$ b.spence@qmw.ac.uk}
\\

\newpage


\section{Introduction}

The subject of D-branes continues to occupy a central role in
current investigations of M-theory in theoretical physics, with
diverse applications.
One such application is in the area of topological quantum field theories, where in the past few years it has begun to be
appreciated that
these models can be understood as arising as effective
theories on the world-volumes of D-branes. 
For example, the twisted $N=4$ super Yang-Mills theories of
\cite{Yamron,VW,Marcus,LabLoz} can be understood in this way
as wrapped D3 branes on calibrated submanifolds of compactification
manifolds of restricted holonomy \cite{BSV}. Various other topological
field theories in two, three and four dimensions
 were discussed from this point of view in
\cite{BT}. Higher dimensional analogues of some of these
models were described in \cite{AOS,BKS,DT,BT2,AFOS}. 
(All of these models should be more properly understood in terms of
the Euclidean \lq\lq E-branes\rq\rq\ of \cite{CMH}.).

Generally, the effective theory on a D-brane world volume is
given in terms of a Born-Infeld theory, giving a non-linear
version of the Maxwell theory, coupled to other fields.
It is thus natural to expect that, when the brane is wrapped
upon manifolds to give effective theories which are topological,
this effective theory ought to be a topological field theory based
upon the Born-Infeld action. We would like to investigate the structure of
such theories
here. It is of course rather difficult to work with these non-linear
models, and we will not present complete actions; however,
the results presented here indicate how to formulate
these theories. In particular, in the pure gauge sector, we will present a Born-Infeld
version of the Abelian Donaldson-Witten theory. 
Then, in the pure embedding sector, where one considers the bosonic fields defining the embedding of the brane in the ambient space, we will discuss
the Born-Infeld analogue of the ampicheiral twisted $N=4$ super Yang-Mills
theory, describing special Lagrangian submanifolds in Calabi-Yau 4-fold compactifications.


\section{A Topological Born-Infeld Theory}

In this section, we will present a topological field theory 
which is a Born-Infeld version of the Abelian Donaldson-Witten
theory. We will first briefly review the latter in order to set the 
notation and motivate and simplify the
subsequent discussion.

\subsection{The Abelian Donaldson-Witten model}

This theory in four dimensions contains the gauge field
$A_i$ ($i=1,..,4$),with field strength $F_{ij}=2\partial_{[i}A_{j]}$,
 an anticommuting vector field
$\psi_i$, an anticommuting self-dual tensor $\chi_{ij}$
($\tilde\chi_{ij} := {1\over2}{\epsilon_{ij}}^{kl}\chi_{kl}
=\chi_{ij}$), an anticommuting scalar field $\eta$, and two
commuting scalar fields $\phi, \lambda$. We will use the
conventions of \cite{Witten}, and will use a condensed
notation, where $F^2=F\cdot F = F^{ij}F_{ij},
\partial^2=\partial^i\partial_i, \partial\psi\cdot\chi = 
\partial_{[i}\psi_{j]}\chi^{ij}, F\tilde F = F\cdot \tilde F = 
F^{ij}\tilde F_{ij}$, etc.

The theory in flat space has the Lagrangian
\be\label{DWflat}
L = {1\over4}F^2 + {1\over2}\phi\partial^2\lambda - i \eta\partial\cdot\psi
+i\partial\psi\cdot\chi + {1\over4}F\tilde F.
\ee
This Lagrangian is invariant under the BRST transformations
\ba\label{DWbrst}
\delta A_j &=& i\psi_j \qquad \delta\psi_j = -\partial_j\phi, \nonumber  \\  
\delta\chi &=& F+\tilde F \qquad \delta\phi = 0, \nonumber  \\ 
\delta\lambda &=& 2i\eta \qquad \delta\eta = 0. 
\ea
These transformations are nilpotent, $\delta^2=0$, up to a gauge transformation
and using the $\chi$ equations of motion $e:= -{i\over2}(\partial\psi + \widetilde{\partial\psi}) = 0$.

This Lagrangian is BRST exact up to the equations of motion $e$:
\be
L = \delta V + {1\over2}\chi\cdot e
\ee
where 
\be
V = {1\over4}F\cdot\chi + {1\over2}\psi\cdot\partial\lambda.
\ee

The model is now defined on a manifold with metric $g_{ij}$. The Lagrangian
becomes $\sqrt g$ times the Lagrangian above, 
with the metric $g_{ij}$ and its inverse used to raise and
lower indices. The BRST symmetry is maintained. To preserve the
self-duality constraint upon $\chi$ on a general manifold, one must
require that $\chi$ transforms under metric variations $\delta g_{ij}$ as
\be\label{chivary}
\delta\chi_{ij} = {1\over2}\epsilon_{ijkl}\delta g ^{km}{\chi_m}^l
                  -\quarter g_{kl}\delta g^{kl}\chi_{ij}.
\ee 
The stress tensor of the model is defined by $T_{ij} = {2\over\sqrt{g}}{\delta S\over\delta g^{ij}}$, where $S$ is the action. 
One finds the expression (with $F_{(i}F_{j)} = {F_{(i}}^kF_{j)k}$,
 $\partial\psi_{i}\chi_{j} = \partial_{[i}\psi_{k]}{\chi_j}^k$)  
\ba
T_{ij} &=& F_{(i}F_{j)} - \quarter g_{ij}F^2 + 2i\partial\psi_{(i}\chi_{j)}
         -\half i \gij\partial\psi\cdot\chi - \partial_{(i}\phi\partial_{j)}
                      \lambda \nonumber \\ 
          && 
        + \half\gij\partial\phi\cdot\partial\lambda + 2i \partial_{(i}\eta\psi_{j)}
    -ig_{ij}\partial\eta\cdot\psi.
 \ea
This is BRST exact
\ba
T_{ij} &=& \delta \Lambda_{ij}, \nonumber \\
\Lambda_{ij} &=& F_{(i}\chi_{j)} - {1\over4}g_{ij}
             F\cdot\chi + \psi_{(i}\partial_{j)}\lambda - 
         {1\over2}g_{ij}\psi\cdot\partial\lambda.
\ea
It is natural to try to relate $\Lambda_{ij}$ directly to $V$. In fact one
has
\be\label{Lambdaeqn}
\Lambda_{ij} = 2{\partial V\over\partial_\chi g^{ij}} - g_{ij} V,
\ee
where the notation ${\partial V\over\partial_\chi g^{ij}}$
means that in varying $V$ by varying the metric, one does
not include the metric variations of the field $\chi$.
We will also use the notation ${\partial V\over\partial_{\rm only} g^{ij}}$
to mean that in varying $V$ by varying the metric, we {\it only} consider
the metric induced variation of $\chi$ given in (\ref{chivary})
above. 

The proof of the relation (\ref{Lambdaeqn}) is as follows:
\ba\label{Tproof}
 T_{ij} &=&  2{\partial L\over\partial g^{ij}} - g_{ij} L \nonumber \\
      &=&  2{\partial \delta V\over\partial g^{ij}} - g_{ij}\delta V
          + \chi_{(i}e_{j)} - \quarter\gij\chi\cdot e\nonumber \\
     &=&   2{\partial \delta V\over\partial_\chi g^{ij}} - g_{ij}\delta V
        + 2{\partial \delta V\over\partial_{\rm only} g^{ij}} 
        + \chi_{(i}e_{j)} - \quarter\gij\chi\cdot e \nonumber \\
    &=&   \delta\left(2{\partial V\over\partial_\chi g^{ij}} - g_{ij} V \right)
          + \chi_{(i}e_{j)} - \quarter\gij\chi\cdot e
               + 2{\partial \over\partial_{\rm only}
                    g^{ij}}\left(\half i\partial\psi\cdot\chi\right)\nonumber \\
  &=&  \delta\left(2{\partial V\over\partial_\chi g^{ij}} - g_{ij} V \right)\nonumber \\
&=& \delta \Lambda_{ij} 
\ea

The fact that the stress tensor is BRST exact implies that correlation
functions of BRST invariant quantities are independent of identity connected
metric variations, and hence define topological invariants.
To calculate these invariants one may use the metric independence
and consider either the weak or strong coupling limit.
In the weak coupling limit, the action localises to its minima. 
This restricts one to the moduli space of instantons, and the invariants
are certain intersection numbers on this moduli space.
Let us now describe a generalisation of this model.

\subsection{A Born-Infeld Topological Theory}

The Born-Infeld Lagrangian in four flat dimensions is given by
(including a total derivative)
\be\label{BIlag}
L = \sqrt{{\det}(1 + F)} -1 + {1\over4}F\cdot\tilde F.
\ee
Now we note the following identity
\be\label{ident1}
\sqrt{{\det}(1 + F)} = 1-{1\over4}F\cdot\tilde F
                      +{1\over4}FF_+\left(1-\half F_-^2 h^2(F)\right),
\ee
where $F_\pm := F \pm \tilde F$, and $h$ is a particular function of
$F$. We have used the fact that this term is positive to write this function as a
square. The above identity is most easily proven by going to a basis in which
\be
F = 
\left(\begin{array}{cccc}
0 & x & 0 & 0\\
-x & 0&0 &0 \\
0&0&0&y \\
0&0&-y&0
\end{array}
\right),
\ee
for some numbers $x, y$. In this basis $\sqrt{{\det}(1 + F)}=\sqrt{(1+x^2)(1+y^2)}$.
Expanding this in powers of $x$ and $y$, we have $\sqrt{(1+x^2)(1+y^2)} = 1 + \half(x^2+y^2) +
g(x,y)$, defining the function $g$. However, $g(x,x)=0$, and $g(x,y)$ is invariant under the
interchange of $x$ and $y$, and also under the replacements $x\rightarrow -x$ or $y\rightarrow -y$.
Thus $(x^2-y^2)^2$ must be a factor of $g$. Thus we can write
\be
\sqrt{(1+x^2)(1+y^2)} = 1 + \half(x^2+y^2) - (x^2-y^2)^2h^2(x,y),
\ee  
defining the function $h$. Note that
$\sqrt{(1+x^2)(1+y^2)} -( 1 + \half(x^2+y^2)) \leq 0$, with equality only if $x=y$.
Re-expressing the equation above in terms of $F$ then yields the identity
(\ref{ident1}).

The Born-Infeld Lagrangian (\ref{BIlag}) 
can then be written as
\ba\label{BIlag2}
L &=& \sqrt{{\det}(1 + F)} -1 + {1\over4}F\cdot\tilde F \nonumber \\
  &=& {1\over8}F_+^2  f^2(F),
\ea
where $f^2 = 1 -{1\over4}F_-^2 h^2$. 
Thus the Born-Infeld action vanishes on instanton solutions, for which $F_+=0$.

Now consider the transformations
{\ba\label{DWBIbrst}
\delta A_j &=& i\psi_j  \qquad \delta\psi_j = -\partial_j\phi \nonumber \\  
\delta\phi &=& 0 \qquad \delta\lambda = 2i\eta \nonumber \\ 
\delta\eta &=& 0 \qquad \delta\chi = F_+f.
\ea}
Suppose that the $\chi$ field equations are $E:= -{1\over4}\delta(F_+f)=0$.
Then the transformations above are nilpotent, up to a gauge transformation,
and using the $\chi$ field equations.
The following Lagrangian contains the Born-Infeld Lagrangian (\ref{BIlag}) and
has the correct $\chi$ field equations to ensure the nilpotence of the
BRST transformations (\ref{DWBIbrst}):
\be
L = {1\over4}F\cdot F_+ f^2 - {1\over2}\chi\cdot\delta(Ff) + {1\over2}\phi\partial^2\lambda - i \eta\partial\cdot\psi.
\ee
Moreover, this Lagrangian is exact up to the field equation $E$:
\be
L = \delta V + {1\over2}\chi\cdot E
\ee
where 
\be
V = {1\over4}(F\cdot\chi)f + {1\over2}\psi\cdot\partial\lambda.
\ee
We now put this theory on a four manifold with metric $g_{ij}$,
as we did with the Abelian Donaldson-Witten theory above, using the
metric $g_{ij}$ to contract indices and including $\sqrt g$ in the measure.
The BRST transformations remain nilpotent on-shell and up to gauge transformations.

Remarkably, the stress tensor of this model is BRST exact :
\ba\label{dipsy}
T_{ij} &=& \delta \Lambda_{ij}  \nonumber \\ 
\Lambda_{ij} &=&  2{\partial V\over\partial_\chi g^{ij}} - g_{ij} V
                   +{1\over2}\chi\cdot F f_{ij}
\ea
($f_{ij}:={\partial f \over \partial g^{ij}}$). 
The proof of (\ref{dipsy}) mirrors that given for the Donaldson-Witten
theory above. Note that $\delta$ and
${\partial\over\partial_\chi g^{ij}}$ do not commute when acting on $V$ here:
\be
\delta {\partial V\over\partial_\chi g^{ij}} - {\partial \delta V\over\partial_\chi g^{ij}} = 
     -{1\over4} F\cdot F_+  ff_{ij}.
\ee
The term on the right-hand side of this equation then combines with another
surviving term to give the last term in $\Lambda_{ij}$ above.
The proof of (\ref{dipsy}) is as follows:
\ba\label{TproofBI}
 T_{ij} &=&  2{\partial L\over\partial g^{ij}} - g_{ij} L \nonumber \\
      &=&  2{\partial \delta V\over\partial g^{ij}} - g_{ij}\delta V
          + \chi_{(i}E_{j)} - \quarter\gij\chi\cdot E\nonumber \\
     &=&   2{\partial \delta V\over\partial_\chi g^{ij}} - g_{ij}\delta V
        + 2{\partial \over\partial_{\rm only} g^{ij}}
          \left(\half i \partial\psi\cdot\chi f - \quarter F\cdot\chi(\delta f)
            \right) 
        + \chi_{(i}E_{j)}  \nonumber \\
   \qquad && - \quarter\gij\chi\cdot E
                - \half \chi^{kl}\delta\cdot(F_{kl}f_{ij})
                 \nonumber \\
    &=&   \delta\left(2{\partial V\over\partial_\chi g^{ij}} - g_{ij} V \right)
          + \half(F\cdot F_+) f f_{ij}
              - \half \chi^{kl}\cdot\delta(F_{kl}f_{ij} )
           \nonumber \\
      &=& \delta\left(2{\partial V\over\partial_\chi g^{ij}} - g_{ij} V
     +\half \chi\cdot F f_{ij}  \right)  
\ea

As with the Donaldson-Witten theory, the
fact that the stress tensor is BRST exact implies that correlation
functions of BRST invariant quantities are independent of metric variations
and hence define topological invariants.
The BRST invariant quantities are the same as for the Donaldson-Witten model.
We also remark that we have not used the explicit form of the function
$f$ in the above discussion, apart from its extremal properties.


\section{Calibrated Submanifolds and Topological Born-Infeld Actions}

We have seen in the previous section that the pure four dimensional
Born-Infeld gauge theory has a topological analogue. 
One can understand why one would expect such a
theory by considering brane wrappings.
It was argued in \cite{BSV} that if one considered wrapping D3 branes on
four dimensional calibrated submanifolds of a seven or eight dimensional space
(with three or two further transverse dimensions), then the effective theories on
the branes are the topological field theories arising from  $N=4$  super Yang-Mills
theory. An immediate reason to suspect this is that the field contents are
correct. The transverse coordinates within the seven or eight dimensional space
are sections of the normal bundle of the world-volume, and can thus be
non-scalar fields in general. Considering the cases where the four manifold
is a special Lagrangian submanifold of a manifold of $SU(4)$ holonomy,
a Cayley submanifold of a $Spin(7)$ eight manifold, and a co-associative
submanifold of a manifold of $G_2$ holonomy, one finds that these
four transverse coordinates give precisely the required fields for the
three twists of the $N=4$ theory.

This analysis suggests that the {\it full} brane effective action, for
such wrappings, should be a Born-Infeld type action which is topological.
If this is the case, then the pure gauge sector should also be topological,
localising on the instanton moduli space. This must be related to
the theory presented above (after including higher-order fermionic
terms in the action, as dictated by supersymmetry).
As further evidence, we must show that the other sectors of the 
Born-Infeld action for these configurations are also of a topological
nature. In particular, if we consider the sector consisting of the
transverse coordinates just discussed, then one expected
feature is that the Born-Infeld actions for these fields should
localise on the space of calibrated submanifolds - ie the
localisation conditions should be the embedding equations for the
calibrated submanifolds.
Relevant properties of these calibrated spaces can be found
in \cite{HL}. In this reference, for the various calibrated
submanifolds, the authors show how to write the square of the volume form
of the submanifold as a sum of squares, with the vanishing of
each expression which is squared being equivalent to the
requirement that the submanifold be calibrated. 

Consider the case of wrapping
a D3 brane on a special Lagrangian submanifold of a Calabi-Yau
four-fold compactification from ten dimensions. The effective Yang-Mills
theory on the brane is then the
\lq\lq ampicheiral\rq\rq\ twisted $N=4$ super Yang-Mills
theory, described in \cite{LabLoz} for example. The full theory
should be a generalisation of this based upon the Born-Infeld theory.
Let us suppose this is the case, and then restrict ourselves
to considering the four transverse fields, which
form a one-form $X_i$.
The relevant Lagrangian is then
\be\label{slagBI}
\sqrt{{\det}(\delta_{ij} + \partial_iX_k\partial_jX^k)} - 1.
\ee 
The embedding equations for the submanifold to be special
Lagrangian are \cite{HL}
\ba\label{slageqns}
&&  \partial_{[i}X_{j]} = 0 \nonumber \\
&&   \partial_i X_i -{\rm det}_{i|i}(\partial X) = 0
\ea
where ${\rm det}_{i|i}$ means the determinant of the matrix with the $i$th row and column removed.

We now wish to see that the Lagrangian (\ref{slagBI}) localises on solutions to
these equations. 
We will define the matrix $H$ to have elements $H_{ij} = \partial_iX_j$.
For any $4\times 4$ matrix $M$, we note the result
\be
\det(1+M) = \sum_{i=0}^4 c_i(M)
\ee  
where
\ba
c_0(M) &=& 1, \qquad c_1(M) = (M) \nonumber \\ 
  c_2(M)&=& \half(M)^2 - \half(M^2) = M^{[i}_{i}M^{j]}_{j}  \nonumber \\
 c_3(M) &=& {1\over3}(M^3) - \half(M)(M^2)+{1\over6}(M)^3=
         M^{[i}_{i}M^{j}_{j}M^{k]}_{k} \nonumber  \\
  c_4(M) &=& \det(M) =  M^{[i}_{i}M^{j}_{j}M_k^kM^{l]}_{l}. 
\ea
We are using a shorthand where $(M)=\tr(M), (M^2)=\tr(M^2)$, etc.
The abbreviation $c_i= c_i(H)$ will be used in the following.

Define
\be
d_\pm = \det(1 \pm iH) = (1-c_2+c_4) \pm i(c_1 - c_3).
\ee
Notice that $c_2$ and $c_4$ are total derivatives:
\ba
  c_2 &=& \partial_i \left (X^{[i} \partial_jX^{j]} \right)\nonumber \\
c_4&=&\partial_i\left( X^{[i} \partial_jX^j\partial_kX^k\partial_lX^{l]} \right). 
\ea

To clarify the issues, we will first simplify the analysis by considering
solutions of the first equation in (\ref{slageqns}).
Then the Hessian matrix $H$ has elements $H_{ij} = \partial_i\partial_jX$,
for some function $X$.
The embedding equations for the submanifold to be special
Lagrangian can be written $c_1-c_3=0$, or equivalently
\be\label{embedding}
d_+-d_- = 0,
\ee
and the Born-Infeld Lagrangian (\ref{slagBI}) is 
\be\label{BIdd}
L = \sqrt{d_+d_-}.
\ee
Now note the identity
\be\label{BIlocal}
\sqrt{d_+d_-} =  (1- c_2 + c_4) -{1\over4}\left(d_+-d_-\right)^2P^2,
\ee
where $P^2$ is a positive function of the $c_i$.
To prove (\ref{BIlocal}), note that since $H_{ij}$ is symmetric, we can diagonalise it
so that $H = {\rm diag}(x_1,x_2,x_3,x_4)$ for some $x_i$. Then
$\sqrt{{\rm det}(1+H^2)} = \sqrt{\prod_{i=1}^4(1+x_i^2)}$. Expanding
this in terms of the symmetric polynomials $c_i$ gives
$(1-c_2+c_4) + g(c_i)$ for some function $g$. However, $g$ must vanish when
$c_1=c_3$. Furthermore, it must also be invariant under the replacements
$c_1\rightarrow -c_1$, $c_3\rightarrow -c_3$. Thus $(c_1-c_3)^2$
factors $g$. The remainder is negative since $\sqrt{d_+d_-}\leq\half(d_++d_-)$, and
vanishes only if $d_+=d_-$.

Thus, from (\ref{BIlocal}), the Born-Infeld Lagrangian (\ref{slagBI}),
when $\partial_{[i}X_{j]} = 0$, can be expressed as a total derivative plus a term
which localises on the embedding equations for the
submanifold to be special Lagrangian, ie $c_1-c_3=0$.
The relevant part of the topological action in this case contains the term
$\delta(\eta(c_1-c_3)P)$, where the fermion $\eta$ has the BRST transformation
$\delta\eta = (c_1-c_3)P$ (this guarantees on-shell nilpotence).

Now consider relaxing the restriction $\partial_{[i}X_{j]} = 0$. 
We seek an expression for
$\sqrt{{\rm det}(1+HH^T)}$, with $H_{ij}=\partial_iX_j$, as a total
derivative plus terms which localise on the solutions of the embedding
equations (\ref{slageqns}).
Split $H_{ij} = \partial_iX_j$ into the sum of a symmetric matrix $S_{ij}$ and
antisymmetric matrix $A_{ij}$, and define $\mu_i = c_i(HH^T+2iA)- c_i(HH^T)$ and $\Delta=-\sum_{i=1}^4\mu_i$. Then we have
\ba\label{detone}
 {\rm det}(1+HH^T) &=& {\rm det}\vert 1+iH\vert^2 - \sum_{i=1}^4\mu_i    \nonumber \\ 
   &=& (1-c_2+c_4)^2 + (c_1-c_3)^2 + \Delta,
\ea
and some calculating yields
\ba\label{3didentity}
\Delta &=& -2(A^2) - 2(S^2)(A^2)+4(S^2A^2)+4(X)(XA^2)-(X)^2(A^2)-6(X^2A^2)
             \nonumber \\
    && +(X^2)(A^2),
\ea
with $X=HH^T=S^2-A^2+[A,S]$. To express $\Delta$ as a sum of squares
minimised by solutions of (\ref{slageqns}), we turn to the results of
\cite{HL} (this can also form a basis for the curved space formulation).
Using their notation, the relevant equation (equation (1.16) of this reference) is
\be\label{HLfour}
 \vert dz^1 dz^2 dz^3 dz^4(\zeta)\vert^2
   + \sum_{(i,j)=1}^4\vert dz^i dz^j k(\zeta)\vert^2
    + \vert {1\over2}k^2(\zeta)\vert^2 = \vert\zeta\vert^2.
\ee
The $z^i$ are complex coordinates on $R^8$, $k$ is the Kahler form 
and $\zeta$ defines a four-plane. To make the connection with the earlier
approach, define $z^i = x^i + iX^i$, so that $dz^i = (1+iH)_{ji}dx^j$.
Then the Kahler form $k = \sum_{i=1}^4 dz^i d\bar z^i$ has components
\be\label{kahler}
 k_{st} = [(1+iH)(1-iH^T)]_{[st]} = 2iA_{st}.
\ee
Then we find the identity
\be\label{bigidentity}
 24\Delta  =     {1\over2}\vert A_{ij}A_{kl}\epsilon^{ijkl}\vert^2
   +   \sum_{(i,j)=1}^4 \vert (1+iH)_{mi}(1+iH)_{nj} A_{pq} \epsilon^{mnpq} \vert^2.
\ee
This is proved by a somewhat laborious comparison with (\ref{3didentity}).
Thus we come to the result, 
\be\label{anotherone}
 {\rm det}(1+HH^T) = \lambda^2 + c^2 + \half\alpha^{ijkl}\alpha_{ijkl} + {1\over6}\beta^{ij}\beta^\dagger_{ij},
\ee
where
\ba\label{moremore}
 \lambda &=& 1-c_2+c_4,     \nonumber \\ 
   c &=&  c_1-c_3, \nonumber \\
  \alpha_{ijkl} &=&  A_{[ij}A_{kl]}  \nonumber \\
  \beta_{ij} &=& \half(1+iH)_{mi}(1+iH)_{nj} A_{pq} \epsilon^{mnpq}.
\ea
Performing an expansion in powers of $\lambda$, one can then write
\be\label{anotherone2}
 \sqrt{{\rm det}(1+HH^T)} = \lambda + c^2A + \alpha^{ijkl}\alpha_{ijkl}B + \beta^{ij}\beta^\dagger_{ij}C,
\ee
for some (positive, if we take the positive square root of $\lambda^2$) functions
$A,B,C$, whose explicit form will not concern us presently.
One can see from this expression that the Lagrangian localises on special
Lagrangian submanifolds (those with $c=\beta_{ij}=\alpha_{ijkl} = 0$).

Turning to the formulation of a topological field theory in this case, consideration of the earlier formulation of the topological Abelian DW theory
leads us to the introduction of anticommuting fermionic fields $\eta, \chi_{ij}$
and $\rho_{ijkl}$ (or $\rho$ with $\rho_{ijkl}=\epsilon_{ijkl}\rho$). The proposed
BRST variations are $\delta\eta = c\sqrt{A}, \delta\chi_{ij} =
\beta_{ij}\sqrt{B}$ and $\delta\rho_{ijkl} = \alpha_{ijkl}\sqrt{C}$, and
the corresponding terms in the Lagrangian are 
$\delta(\eta c\sqrt{A} + \chi^{ij}\beta_{ij}\sqrt{B} + \rho^{ijkl}\alpha_{ijkl}\sqrt{C})$.
Examination of the stress tensor suggests that it is exact - for example, in the
sector of the theory containing the field $\eta$, we have
$\delta\eta = \partial^iX_i, \delta X_i = \tilde\psi_i, \delta \tilde\psi_i=0$.
With the Lagrangian $L = \delta(\eta\partial^iX_i) - \eta\delta(\partial^iX_i)$,
up to equation of motion terms one finds that the stress tensor is given by
$T_{ij} = \delta(4\eta\partial_{(i}X_{j)})$.


\section{Remarks}

 We have proposed in this letter that
there are topological field theories
based upon Born-Infeld theories, which generalise the known twisted
supersymmetric Abelian gauge theories. For the twisted $N=2$ case, we have
given a Born-Infeld generalisation of the Abelian Donaldson-Witten theory and
shown explicitly that it is topological.
For the Born-Infeld models corresponding to the twisted
$N=4$ Abelian theories, we have considered the conjecture that these are the 
full effective theories of D3 branes wrapped on certain calibrated submanifolds, for the case of special Lagrangian four-manifolds.
This was supported by the localisation properties of the action for this case,
although the full result is not yet apparent.
This work suggests the general result in the $N=4$ case
that the Born-Infeld actions for the transverse fields localise on the
moduli space of calibrated submanifolds and yield the relevant
topological field theories. An obvious candidate for the full theory, including the
higher order fermion terms, is the
supersymmetric Born-Infeld theory of \cite{Schwarz}, defined on the appropriate
spaces. World-sheet supersymmetry becomes BRST invariance, and as already
analysed in \cite{GLW}, the conditions for supersymmetry are the embedding equations
for the calibrated submanifolds. 
One would also expect a Born-Infeld version of the DW theory which is non-linear in
the fermions, from a similar source. Analogues of this in higher dimensions,
being Born-Infeld versions of the theories in \cite{AOS,BKS,DT,BT2,AFOS}, should also
exist, being the theories describing wrapped Euclidean D-branes in these
cases. The rather simple structure which is emerging in this work suggests that
these topological theories may have a relatively simple formulation.

\vspace{2cm}

{\centerline {\bf Acknowledgements}}
This work was supported by
 an EPSRC Advanced Fellowship. I would like to thank Jos\'e Figueroa-O'Farrill for helpful conversations.

\vspace{1cm}

{\centerline {\bf Note Added}}
As this work was being written up, the paper \cite{Imaanpur} appeared,
which discusses a topological field theory which localises on special
Lagrangian submanifolds of a Calabi-Yau three fold. This must be related to
a three dimensional analogue of the four dimensional model discussed
above.

\end{document}